\documentclass[conference]{IEEEtran}
\usepackage{amssymb}
\usepackage{bbold}
\usepackage{bbm}
\usepackage[cmex10]{amsmath}
\usepackage{stfloats}
\usepackage{graphicx}
\usepackage{multirow}
\usepackage{subfigure}
\usepackage{tabularx}
\usepackage{booktabs}
\usepackage{amsmath}
\usepackage{epsfig,epsf,color,balance,cite}
\usepackage{indentfirst}
\usepackage{mathrsfs}
\usepackage{algorithm}
\usepackage{algorithmic}
\usepackage{dsfont}

\begin{document}
\title{Joint Beamforming and Computation Offloading for Multi-user Mobile-Edge Computing}
\author{
\IEEEauthorblockN{ Changfeng Ding\IEEEauthorrefmark{1},
                  Jun-Bo Wang\IEEEauthorrefmark{1}\IEEEauthorrefmark{2},
                  Ming Cheng\IEEEauthorrefmark{1},
                  Chuanwen Chang\IEEEauthorrefmark{3},
                  Jin-Yuan Wang\IEEEauthorrefmark{4}, and
                  Min Lin\IEEEauthorrefmark{4}
                  }
\IEEEauthorblockA{\IEEEauthorrefmark{1}National Mobile Communications Research Laboratory, Southeast University, Nanjing 211111, China.}
\IEEEauthorblockA{\IEEEauthorrefmark{2}School of Cyber Science and Engineering, Southeast University, Nanjing 211111, China.}
\IEEEauthorblockA{\IEEEauthorrefmark{3}The 28th Research Institute of China Electronics Technology Group Corporation, Nanjing 210007, China.}
 \IEEEauthorblockA{\IEEEauthorrefmark{4}School of Science, Nanjing University of Posts and Telecommunications, Nanjing 210003, China.}
\IEEEauthorblockA{E-mail: \{cfding, jbwang, mingcheng\}@seu.edu.cn, cwchang28@foxmail.com, \{jywang, linmin\}@njupt.edu.cn}
}

\maketitle
\begin{abstract}
Mobile edge computing (MEC) is considered as an efficient method to relieve the computation burden of mobile devices. In order to reduce the energy consumption and time delay of mobile devices (MDs) in MEC, multiple users multiple input and multiple output (MU-MIMO) communications is considered to be applied to the MEC system. The purpose of this paper is to minimize the weighted sum of energy consumption and time delay of MDs by jointly considering the offloading decision and MU-MIMO beamforming problems. And the resulting optimization problem is a mixed-integer non-linear programming problem, which is NP-hard. To solve the optimization problem, a semidefinite relaxation based algorithm is proposed to solve the offloading decision problem. Then, the MU-MIMO beamforming design problem is handled with a newly proposed fractional programming method. Simulation results show that the proposed algorithms can effectively reduce the energy consumption and time delay of the computation offloading.
\end{abstract}

\IEEEpeerreviewmaketitle

\section{Introduction}
\label{section1}
Based on the fast development of electronic devices and mobile communication technologies, smart mobile devices (MDs) are expected to run more complicate applications such as face recognition, interactive gaming, and augmented reality, etc. Unfortunately, due to the constraints of cost and physical size, most MDs do not have enough energy supply or computation capability to run complicate applications. Therefore, efficient methods are required to relieve the contradiction between complicate applications' demands and limited resources at MDs.

Driven by the growing demands of Internet of Things and mobile applications, mobile edge computing (MEC) is considered as a promising way to enable computation-intensive and latency-critical applications at the resource-limited MDs \cite{MEC01}. In MEC, the computing server is implemented at the edge of the network. Therefore, the computation intensive tasks of MDs can be handled by those edge node to reduce the energy consumption and time delay of certain applications. To promote the application of MEC, a lot of work has been done. The computation offloading strategy and radio resources allocation are jointly considered in an edge and central cloud computing system to optimize the energy consumption and time delay in \cite{MEC02}. The authors in \cite{MEC03} studied an offloading system where a MD can offload computation tasks to multiple edge servers, in which fixed and elastic CPU frequency are considered for MD. In most of the existing studies, MDs and base station (BS) are assumed to be equipped with single antenna, only a few researchers have applied the multi-antenna technology to MD or BS. Multi-antenna based energy harvesting strategy was introduced into MEC system \cite{MEC04} to enhance the computation capability and prolong the operation time of MDs. To realize wireless backhaul and exploit benefits from multi-antenna, the authors in \cite{MEC05} proposed to use receive beamforming at a multi-antenna BS, while multiple single antenna MDs transmit by using orthogonal multiple access techniques such as time/frequency-division. To improve the computation offloading efficiency, the authors in \cite{MEC06} utilized multi-antenna at both MD and BS to realize MIMO transmission. However, the MDs in the small cell still use orthogonal frequency resources.

With the growing of communication technologies, some novel techniques, such as massive MIMO technology \cite{MEC07}, can be adopted to improve system's performance. Uplink MU-MIMO communications was studied in \cite{MEC08}, and the results show that uplink MU-MIMO communications can lead to significant improvement in cell throughput. The work in \cite{MEC09} showed that using multi-antenna for MIMO transmission can effectively reduce the transmission power and increase energy efficiency. Inspired by this, applying MU-MIMO transmission to MEC may further bring extra benefits and improve the performance of energy consumption and time delay during computation offloading.

In this paper, we consider the fusion of MEC and MU-MIMO communication to realize the MU-MIMO computation offloading for multiple MDs in the coverage area of a BS. The optimization problem is formulated as a joint offloading decision-making and MU-MIMO beamforming design problem under the constraints of maximum transmit power and time delay. The problem is a mixed-integer non-linear programming problem, which is NP-hard. Due to the complexity of the optimization problem, we deal with the optimization problem in two steps. First, the optimization problem is reformulated via quadratic constrained quadratic programming (QCQP) and semidefinite relaxation (SDR). And the offloading decisions are obtained through the approximation of the SDR solution as in Algorithm 1. Then, a fractional programming based method is adopted to transform the non-convex MU-MIMO beamforming problem into a convex one. And the optimal beamforming matrices are acquired via the proposed Algorithm 2. Simulation results show that the proposed MU-MIMO computation offloading method can effectively reduce the time delay and energy consumption.

\section{System Model}
\label{section2}
\begin{figure}
\centering
\includegraphics[width=6.5cm]{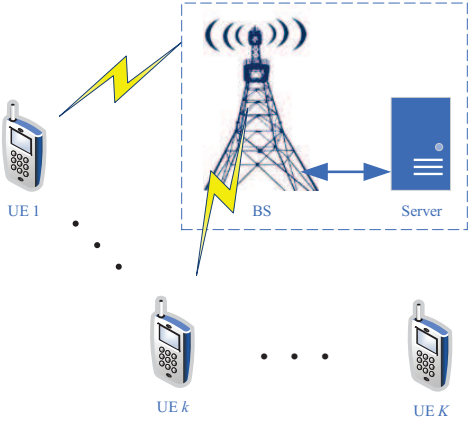}
\caption{Illustration of a MEC system}
\label{fig1}
\end{figure}

As shown in figure \ref{fig1}, this paper considers a MEC system that consists of a BS equipped with a MEC server and $M$ antennas and $K$ MDs equipped with $N$ antennas. Define ${\mathds{U}} = \left\{ {1, \ldots ,K} \right\}$ as the set of MDs.
Then for MD $k$ ($k \in {\mathds{U}}$), its computation task can be processed locally or offloaded to the BS. With the collected information from MDs (e.g. computation task size, local computing capability, channel state information and maximum tolerable delay), the BS can make optimized offloading decisions for MDs. A quasi-static scenario is considered where all the MDs remain stationary during an offloading period \cite{MEC03}.

In this paper, uplink MU-MIMO communications are considered for computation offloading. Without loss of generality, each MD supports the transmission of $d$ streams over $N$ antennas ($d \le N$ and $Kd \le M$). Let ${{\mathds{U}}_o} = \left\{ {1, \ldots ,{N_o}} \right\} \subseteq {\mathds{U}}$ be the set of offloading MDs. When MD $k$ ($k \in {{\mathds{U}}_o}$) offloads its task to the BS, the received signal of ${l_{th}}$ data stream from MD $k$ at BS can be expressed as \cite{MEC09}
\begin{small}
\begin{eqnarray}
{y_{k,l}} &=& {\bf{v}}_{k(l)}^H{{\bf{H}}_k}{{\bf{q}}_{k(l)}}{x_{k,l}} \nonumber \\
&+& {\bf{v}}_{k(l)}^H\sum\limits_{i \in {{\mathds{U}}_o},i \ne k} {\sum\limits_{j = 1}^d {{{\bf{H}}_i}{{\bf{q}}_{i(j)}}{x_{i,j}}} }  + {\bf{v}}_{k(l)}^H{\bf{n}}
\label{equ1}
\end{eqnarray}
\end{small}where ${{\bf{q}}_{k(l)}}$ and ${{\bf{v}}_{k(l)}}$ are the ${l_{th}}$ column of $N \times d$ transmit beamforming matrix ${{\bf{Q}}_k}$ and $M \times d$ receive beamforming matrix ${{\bf{V}}_k}$, respectively. ${x_{k,l}}$ is the ${l_{th}}$ symbol of the transmitted $d \times 1$ data vector ${{\bf{x}}_k}$, and ${{\bf{x}}_k}$ satisfies $E[{{\bf{x}}_k}{\bf{x}}_k^H] = {I_d}$ and $E[{{\bf{x}}_k}{\bf{x}}_l^H] = {\bf{0}}$ for $k \ne l$. ${{\bf{H}}_k}$ denotes the $M \times N$ channel matrix from MD $k$ to BS. ${\bf{n}}$ denotes the additive white Gaussian noise (AWGN) vector and $E[{\bf{n}}{{\bf{n}}^H}] = {\sigma ^2}{{\bf{I}}_M}$. According to (\ref{equ1}), the achievable data rate ${R_k}$ of the MD ${k}$ can be given by
\begin{small}
\begin{equation}
{R_k} = \sum\limits_{l = 1}^{\rm{d}} {B_W{{\log }_2}\left( {1 + {\bf{q}}_{k(l)}^H{\bf{H}}_k^H{{\bf{v}}_{k(l)}}{I_N^{- 1}}{\bf{v}}_{k(l)}^H{{\bf{H}}_k}{{\bf{q}}_{k(l)}}} \right)}
\label{equ2}
\end{equation}
\end{small}where $B_W$ (Hz) is the system bandwidth and $I_N =  \sum\nolimits_{i \in {U_o},i \ne k} {\sum\nolimits_{j = 1}^d {{\bf{v}}_{k(l)}^H{{\bf{H}}_i}{{\bf{q}}_{i(j)}}{\bf{q}}_{i(j)}^H{\bf{H}}_i^H{{\bf{v}}_{k(l)}}} }  + {\sigma ^2}{\bf{v}}_{k(l)}^H{{\bf{v}}_{k(l)}}$.

The computation task of MD $k$ is presented as ${J_k} = \left( {{B_k},\tau _{\max }^k} \right)$ ($k \in {\mathds{U}}$), where ${B_k}$ (in bits) denotes the size of input data and $\tau _{\max }^k$ is the maximum tolerable delay (in seconds). Since the computation result is usually small, the time delay of receiving computation result is uauslly omitted \cite{MEC10}. For MD $k$ ($k \in {{\mathds{U}}_l} = {\mathds{U}}\backslash {{\mathds{U}}_o}$) in local computing, the local computation time can be expressed as ${T_{loc,k}} = {{{C_k}}}/{{{f_{loc,k}}}}$, where ${C_k} = \alpha {B_k}$ is the total number of CPU cycles to finish task computing and $\alpha $ (cycles/bit) is the processing density, and ${f_{loc,k}}$  is the local computation capability (cycles/s) of MD $k$. According to \cite{MEC11}, the energy consumption of MD $k$ in local computing can be given as ${E_{loc,k}} = \kappa {C_k}f_{loc,k}^2$, where $\kappa $ is a constant related to the hardware architecture \cite{MEC12}.

To simplify the analysis, it is assumed that the BS starts task computation after receiving all MDs' tasks. Hence, the data transmission delay of MD $k$ in ${{\mathds{U}}_o}$ is given as ${T_{tran,k}} = {\max _{i \in {U_o}}}\left\{ {{B_i}/{R_i}} \right\}$. While the execution time delay at BS can be formulated as ${T_{c,k}} = {{{C_k}}}/{{{f_{c,k}}}}$, where ${f_{c,k}}$ (cycles/s) is the computation resource allocated to MD $k$.

According to the analysis stated above, when offloading decision is considered, the time delay of MD $k$ in ${\mathds{U}}$ during computation offloading can be denoted as ${T_{off,k}} = {\max _{i \in U}}\left\{ {{{{c_i}{B_i}} \mathord{\left/{\vphantom {{{c_i}{B_i}} {{R_i}}}} \right.\kern-\nulldelimiterspace} {{R_i}}}} \right\} + {T_{c,k}}$, where ${c_i} \in \left\{ {0,1} \right\}$ is the offloading decision of MD $i$. If ${c_i} = 0$,  the MD $i$ computes its task at local; otherwise, the task is offloaded to the BS.

The circuit energy consumption of MD $k$ during idle time ${T_{c,k}}$ can be formulated as ${E_{c,k}} = p_k^{id}{T_{c,k}}$, where $p_k^{id}$ is the power consumption (in watt) in idle state. Therefore, the energy consumption of MD $k$ during computation offloading can be given as ${E_{off,k}} = \left( {{{{B_k}} \mathord{\left/{\vphantom {{{B_k}} {{R_k}}}} \right.\kern-\nulldelimiterspace} {{R_k}}}} \right)p_k^t + {E_{c,k}}$, where $p_k^t = \left\| {{{\bf{Q}}_k}} \right\|_F^2$ is the transmission power of MD $k$. Since the computing result is usually small, the energy consumption of receiving the computing result is ignored at MD side.
Finally, the time delay and energy consumption of MD $k$ can be computed respectively as
\begin{eqnarray}
{E_k} &=& \left( {1 - {c_k}} \right){E_{loc,k}} + {c_k}{E_{off,k}} \label{equ3}\\
{T_k} &=& \left( {1 - {c_k}} \right){T_{loc,k}} + {c_k}{T_{off,k}} \label{equ4}.
\end{eqnarray}

Based on the above analysis, the objective of this paper is to minimize the weighted sum of energy consumption and time delay of all MDs. Mathematically, the optimization problem is formulated as
\begin{small}
\begin{subequations}
\label{EqObj1}
\begin{eqnarray}
    &P1:{\rm{  }}\mathop {\min }\limits_{{\bf{c}},{\mathds{Q,V}}} {\rm{ }}\sum\limits_{k = 1}^K {\left( {\lambda _k^e{E_k} + \lambda _k^t{T_k}} \right)} \label{EqObj1A}  \\
    &s.t. ~~ {c_k} \in \left\{ {0,1} \right\},\forall k \in {\mathds{U}} \label{EqObj1B}\\
    &p_k^t \le P_{\max }^k{\rm{,}}\forall k \in {\mathds{U}} \label{EqObj1C}\\
    &{T_k} \le \tau _{\max }^k{\rm{,}}\forall k \in {\mathds{U}} \label{EqObj1D}
\end{eqnarray}
\end{subequations}
\end{small}where ${\mathds{Q}} = \left\{ {{{\bf{Q}}_1}, \ldots ,{{\bf{Q}}_K}} \right\}$ and ${\mathds{V}} = \left\{ {{{\bf{V}}_1}, \ldots ,{{\bf{V}}_K}} \right\}$ are the sets of transmit beamforming matrices and receive beamforming matrices, respectively. ${\bf{c}} = {[{c_1},{c_2}, \ldots ,{c_K}]^T}$ is the offloading decisions of all MDs. $P_{\max }^k$ is the maximum transmit power of MD $k$. Note that in (\ref{EqObj1}), the constraint (\ref{EqObj1B}) guarantees that the offloading decision of each MD is binary variable. Constraint (\ref{EqObj1C}) denotes the maximum transmission power of each MD, and constraint (\ref{EqObj1D}) indicates the maximum tolerable delay. It can be observed that (\ref{EqObj1}) is a mixed-integer non-linear programming problem, which is generally NP-hard \cite{MEC13}. Due to space limitation, the proof of the NP-hard problem (\ref{EqObj1}) is omitted.

\section{Uplink MU-MIMO Offloading Decision Making}
\label{section4}
In this section, the QCQP and SDR based transformations are proposed to solve the offloading decision problem. Let $\max \left\{ {{{{B_k}}}/{{{R_k}}}p_k^t} \right\} = p_k^{com}$ and ${\max _{k \in {\mathds{U}}}}\left\{ {{c_k}{{{B_k}}}/{{{R_k}}}} \right\} = t$, we have ${B_k}p_k^t \le p_k^{com}{R_k},  \forall k \in {\mathds{U}}$ and ${B_k}{c_k} \le t{R_k},  \forall k \in {\mathds{U}}$. Then, the problem (\ref{EqObj1}) can be transformed to the problem (\ref{EqObj2}) as
\begin{small}
\begin{subequations}
\label{EqObj2}
\begin{eqnarray}
& P2:{\!\!\!\!\!\!\!\!}\mathop {\min }\limits_{{\bf{c}},{\mathds{Q,V}},{\bf{p}}^{com},t} {\rm{ }}\sum\limits_{k = 1}^K {\left( {{\delta _k}{c_k} \!+\! \lambda _k^e{c_k}p_k^{com} \!+\! \lambda _k^t{c_k}t} \right)}\!+\!\eta \label{EqObj2A} \\
& s.t.~~{\rm{  }} {c_k}\left( {1 - {c_k}} \right) = 0{\rm{  }}, \forall k \in {\mathds{U}} \label{EqObj2B} \\
& {B_k}p_k^t \le p_k^{com}{R_k}{\rm{   }}, \forall k \in {\mathds{U}} \label{EqObj2C} \\
& {B_k}{c_k} \le t{R_k}{\rm{  }}, \forall k \in {\mathds{U}} \label{EqObj2D} \\
& \left( {1 - {c_k}} \right)T_{loc,k} + {c_k}\left( {t + T_{c,k}} \right) \le \tau _{\max }^k{\rm{  }}, \forall k \in {\mathds{U}} \label{EqObj2E} \\
& p_k^t \le P_{\max }^k{\rm{  }}, \forall k \in {\mathds{U}} \label{EqObj2F}
\end{eqnarray}
\end{subequations}
\end{small}where ${{\bf{p}}^{com}} = \left\{ {p_1^{com}, \ldots ,p_K^{com}} \right\}$, ${\delta _k} = \lambda _k^eE_{c,k} + \lambda _k^tT_{c,k}{\rm{ - }}\left( {\lambda _k^eE_{loc,k} + \lambda _k^tT_{loc,k}} \right)$ ($\forall k \in {\mathds{U}}$) and $\eta {\rm{ = }}\sum\nolimits_{k = 1}^K {\left( {\lambda _k^eE_{loc,k} + \lambda _k^tT_{loc,k}} \right)} $ are constant, and $\eta $ can be omitted in the following analysis. The integer constraint in (\ref{EqObj1B}) is replaced by quadratic constraint in (\ref{EqObj2B}).
To perform the QCQP transformation, a $(4K + 1) \times 1$ vector {\bf{s}} is defined as
\begin{small}
\begin{equation}
{\bf{s}} \!=\! {\left[ \!{{c_1}, \!\ldots\!,{c_K},{R_1}, \!\ldots\!,{R_K},p_1^{com}, \!\ldots \!,p_K^{com},p_1^t, \!\ldots \!,p_K^t,t} \right]^T}.
\label{equ7}
\end{equation}
\end{small}

Mathematically, P2 is transformed into the following standard QCQP problem
\begin{small}
\begin{subequations}
\label{EqObj3}
\begin{eqnarray}
& P3:{\rm{  }}\mathop {\min }\limits_s \left( {{{\bf{s}}^T}{{\bf{M}}_1}{\bf{s}} + 2{\bf{c}}_0^T{\bf{s}}} \right) \label{EqObj3A} \\
& s.t.~~{\rm{   }}{{\bf{s}}^T}{\rm{diag}}\left( {{{\bf{e}}_k}} \right){\bf{s}} - {\bf{e}}_k^T{\bf{s}} = 0,{\rm{  }}\forall k \in {\mathds{U}} \label{EqObj3B} \\
& {{\bf{s}}^T}{{\bf{M}}_2}{\bf{s}} + {B_k}\left( {{\bf{e}}_{3K + k}^T{\bf{s}}} \right) \le 0,{\rm{  }}\forall k \in {\mathds{U}} \label{EqObj3C} \\
& {{\bf{s}}^T}{{\bf{M}}_3}{\bf{s}} + \left( {{B_k}{\bf{e}}_k^T} \right){\bf{s}} \le 0,{\rm{  }}\forall k \in {\mathds{U}} \label{EqObj3D} \\
& {{\bf{s}}^T}{{\bf{M}}_4}{\bf{s}} + \left( { - T_{loc,k}{\bf{e}}_k^T} \right){\bf{s}} + T_{loc,k} \le \tau _{\max }^k{\rm{,  }}\forall k \in {\mathds{U}} \label{EqObj3E} \\
& \left( {{\bf{e}}_{3K + k}^T{\bf{s}}} \right) \le P_{\max }^k{\rm{, }}\forall k \in {\mathds{U}} \label{EqObj3F}
\end{eqnarray}
\end{subequations}
\end{small}where ${{\bf{c}}_0} = \left( {1/2} \right){\left[ {{\delta _1}, \ldots ,{\delta _K},{{\bf{0}}_{1 \times 3K}},{0_{1 \times 1}}} \right]^T}$ is a constant $\left( {4K + 1} \right) \times 1$ vector; ${{\bf{e}}_k}$ and ${{\bf{e}}'_k}$ are standard unit vector with size of ${\left( {4K + 1} \right) \times 1}$ and $K \times 1$, respectively. And the matrices of ${{\bf{M}}_1}$, ${{\bf{M}}_2}$, ${{\bf{M}}_3}$, ${{\bf{M}}_4}$, ${{\bf{M}}_5}$, ${{\bf{D}}_e}$, ${{\bf{c}}_t}$ and ${{\bf{D}}_p}$ in (\ref{EqObj3}) are listed as follows
\begin{small}
\begin{equation}
{{\bf{M}}_1} \!=\! \!\frac{1}{2}\!{\left[
\!\!\!\!\!{\begin{array}{*{20}{c}}
{{{\bf{0}}_{K \times K}}}&\!\!\!\!\!\!{{{\bf{0}}_{K \times K}}}&\!\!\!\!\!\!\!{{{\bf{D}}_e}}&{\begin{array}{*{20}{c}}
\!\!\!\!\!\!\!\!\!\!\!\!\!\!\!\!\!\!\!\!{{{\bf{0}}_{K \times K}}}&{{{\bf{c}}_t}}
\end{array}}\\
{{{\bf{0}}_{K \times K}}}&\!\!\!\!\!\!\!{{{\bf{0}}_{K \times K}}}&\!\!\!\!\!{{{\bf{0}}_{K \times K}}}&
\!\!\!\!\!\!\!\!\!{\begin{array}{*{20}{c}}
\!\!{{{\bf{0}}_{K \times K}}}&\!{{{\bf{0}}_{K \times 1}}}
\end{array}}\\
\!{{{\bf{D}}_e}}& \!\!\!\!\!\!\!{{{\bf{0}}_{K \times K}}}&\!\!\!\!\!{{{\bf{0}}_{K \times K}}}&\!\!\!\!\!\!\!\!\!\!{\begin{array}{*{20}{c}}
\!{{{\bf{0}}_{K \times K}}}&\!{{{\bf{0}}_{K \times 1}}}
\end{array}}\\

\!{\begin{array}{*{20}{c}}
{{{\bf{0}}_{K \times K}}}\\
{{\bf{c}}_t^T}
\end{array}}&

\!\!\!\!\!\!\!\!\!{\begin{array}{*{20}{c}}
{{{\bf{0}}_{K \times K}}}\!\!\\
{{{\bf{0}}_{1 \times K}}}
\end{array}}&

\!\!\!\!\!{\begin{array}{*{20}{c}}
{{{\bf{0}}_{K \times K}}}\\
\!{{{\bf{0}}_{1 \times K}}}
\end{array}}&

\!\!\!\!\!\!\!\!\!\!\!{\begin{array}{*{20}{c}}
{\begin{array}{*{20}{c}}
{{{\bf{0}}_{K \times K}}}\\
\!\!{{{\bf{0}}_{1 \times K}}}
\end{array}}&

\!\!\!\!\!\!\!\!{\begin{array}{*{20}{c}}
{{{\bf{0}}_{K \times 1}}}\\
\!\!{{0_{1 \times 1}}}
\end{array}}
\end{array}}
\end{array}} \!\!\!\!\!\!\!\!\!\right]_{\!\!\left( \!{4K \!+ \!1} \right) \!\times\! \left( \!{4K \!+\! 1} \right)}}
\label{equ9}
\end{equation}
\end{small}
\begin{small}
\begin{equation}
{{\bf{M}}_2} \!=\!\! {\left[ \!\!\!\!{\begin{array}{*{20}{c}}
{{{\bf{0}}_{K \!\times\! K}}}&{{{\bf{0}}_{K \!\times\! 2K}}}&{{{\bf{0}}_{K \!\times\! \left( \!{K + 1} \right)}}}\\
{{{\bf{0}}_{2K \!\times\! K}}}&{{{\bf{D}}_p}}&{{{\bf{0}}_{2K \!\times\! \left( \!{K + 1} \right)}}}\\
{{{\bf{0}}_{\left( \!{K + 1} \right) \!\times\! K}}}&{{{\bf{0}}_{\left( \!{K + 1} \right) \!\times\! 2K}}}&{{{\bf{0}}_{\left( \!{K + 1} \right) \!\times\! \left( \!{K + 1} \right)}}}
\end{array}} \!\!\!\!\right]_{\!\!\left( \!{4K \!+\! 1} \right) \!\times\! \left( \!{4K \!+\! 1} \right)}}
\label{equ10}
\end{equation}
\end{small}
\begin{small}
\begin{equation}
{{\bf{M}}_3} \!=\!\!  - \frac{1}{2}\!{\left[ \!\!\!\!{\begin{array}{*{20}{c}}
{{{\bf{0}}_{K \times K}}}&\!\!\!\!{{{\bf{0}}_{K \times K}}}&\!\!\!\!{{{\bf{0}}_{K \times 2K}}}&\!\!\!\!{{{\bf{0}}_{K \times 1}}}\\
{{{\bf{0}}_{K \times K}}}&\!\!\!\!{{{\bf{0}}_{K \times K}}}&\!\!\!\!{{{\bf{0}}_{K \times 2K}}}&\!\!\!\!{{\bf{e}}'_p}\\
{{{\bf{0}}_{2K \times K}}}&\!\!\!\!{{{\bf{0}}_{2K \times K}}}&\!\!\!\!{{{\bf{0}}_{2K \times 2K}}}&\!\!\!\!{{{\bf{0}}_{2K \times 1}}}\\
{{{\bf{0}}_{1 \times K}}}&\!\!\!\!{{{\left( {{\bf{e}}'_p} \right)}^T}}&\!\!\!\!{{{\bf{0}}_{1 \times 2K}}}&\!\!\!\!{{0_{1 \times 1}}}
\end{array}} \!\!\!\!\right]_{\!\left( \!{4K \!+\! 1} \right) \!\times\! \left( \!{4K \!+\! 1} \right)}}
\label{equ11}
\end{equation}
\end{small}
\begin{small}
\begin{equation}
{{\bf{M}}_4} \!=\!\! \frac{1}{2}{\left[ \!\!\!\!{\begin{array}{*{20}{c}}
{{{\bf{0}}_{K \times K}}}&{{{\bf{0}}_{K \times 3K}}}&{{\bf{e}}'_k}\\
{{{\bf{0}}_{3K \times K}}}&{{{\bf{0}}_{3K \times 3K}}}&{{{\bf{0}}_{3K \times 1}}}\\
{{{\left( {{\bf{e}}'_k} \right)}^T}}&{{{\bf{0}}_{1 \times 3K}}}&{{0_{1 \times 1}}}
\end{array}} \!\!\!\!\right]_{\!\left( \!{4K \!+\! 1} \right) \!\times\! \left( \!{4K \!+\! 1} \right)}}
\label{equ12}
\end{equation}
\end{small}
\begin{small}
\begin{equation}
{{\bf{D}}_e} = {\left[ {\begin{array}{*{20}{c}}
{\lambda _1^e}& \cdots &0\\
 \vdots & \ddots & \vdots \\
0& \cdots &{\lambda _K^e}
\end{array}} \right]_{K \times K}}
\label{equ13}
\end{equation}
\end{small}
\begin{equation}
{{\bf{c}}_t} = {\left[ {\lambda _1^t, \ldots ,\lambda _K^t} \right]^T}
\label{equ14}
\end{equation}
\begin{equation}
{{\bf{D}}_p} = \left( { - \frac{1}{2}} \right){\left[ {\begin{array}{*{20}{c}}
{{{\bf{0}}_{K \times K}}}&{diag\left( {{\bf{e}}'_p} \right)}\\
{diag\left( {{\bf{e}}'_p} \right)}&{{{\bf{0}}_{K \times K}}}
\end{array}} \right]_{2K \times 2K}}
\label{equ15}.
\end{equation}

To solve the QCQP problem (\ref{EqObj3}), the SDR method proposed in \cite{MEC14} can be used to transform P3 into a Semidefinite programming (SDP) problem. Define matrix ${\bf{G}} = \left[ {\begin{array}{*{20}{c}}{\bf{s}}\\1\end{array}} \right]\left[ {\begin{array}{*{20}{c}}{{{\bf{s}}^T}}&1\end{array}} \right]$, then, P3 can be rewritten as
\begin{small}
\begin{subequations}
\label{EqObj4}
\begin{eqnarray}
& P4:~~\mathop {\min }\limits_{\bf{G}} {\rm{ }} {\rm{Tr}}\left( {{{\bf{W}}_0}{\bf{G}}} \right) \label{EqObj4A} \\
& s.t.~~{\rm{Tr}}\left( {{{\bf{W}}_1}{\bf{G}}} \right) = 0{\rm{  }}, \forall k \in {\mathds{U}} \label{EqObj4B} \\
& {\rm{Tr}}\left( {{{\bf{W}}_2}{\bf{G}}} \right) \le 0{\rm{  }}, \forall k \in {\mathds{U}} \label{EqObj4C} \\
& {\rm{Tr}}\left( {{{\bf{W}}_3}{\bf{G}}} \right) \le 0{\rm{  }}, \forall k \in {\mathds{U}} \label{EqObj4D} \\
& {\rm{Tr}}\left( {{{\bf{W}}_4}{\bf{G}}} \right) + T_{loc,k} - \tau _{\max }^k \le 0{\rm{  }}, \forall k \in {\mathds{U}} \label{EqObj4E} \\
& {\rm{Tr}}\left( {{{\bf{W}}_5}{\bf{G}}} \right) - P_{\max }^k \le 0{\rm{  }}, \forall k \in {\mathds{U}} \label{EqObj4F} \\
& {\bf{G}} \succeq 0 \label{EqObj4G} \\
& {{\bf{G}}_{\left( {4K + 2} \right) \times \left( {4K + 2} \right)}} = 1 \label{EqObj4H} \\
& {\rm{rank}}\left( {\bf{G}} \right) = 1 \label{EqObj4I}
\end{eqnarray}
\end{subequations}
\end{small}where the matrices of ${{\bf{W}}_0}$, ${{\bf{W}}_1}$, ${{\bf{W}}_2}$, ${{\bf{W}}_3}$, ${{\bf{W}}_4}$, ${{\bf{W}}_5}$ are listed as follows
\begin{small}
\begin{equation}
{{\bf{W}}_0} = {\left[ {\begin{array}{*{20}{c}}
{{{\bf{M}}_1}}&{{{\bf{c}}_0}}\\
{{\bf{c}}_0^T}&0
\end{array}} \right]_{\left( {4K + 2} \right) \times \left( {4K + 2} \right)}}
\label{equ17}
\end{equation}
\end{small}
\begin{small}
\begin{equation}
{{\bf{W}}_1} = {\left[ {\begin{array}{*{20}{c}}
{{\rm{diag}}\left( {{{\bf{e}}_k}} \right)}&{ - 1/2{{\bf{e}}_k}}\\
{ - 1/2{\bf{e}}_k^T}&0
\end{array}} \right]_{\left( {4K + 2} \right) \times \left( {4K + 2} \right)}}
\label{equ18}
\end{equation}
\end{small}
\begin{small}
\begin{equation}
{{\bf{W}}_2} = {\left[ \!\!{\begin{array}{*{20}{c}}
{{{\bf{M}}_2}}&{1/2{B_k}{{\bf{e}}_{3K + k}}}\\
{1/2{B_k}{\bf{e}}_{3K + k}^T}&0
\end{array}} \!\!\right]_{\!\left( \!{4K \!+\! 2} \right) \!\times\! \left( \!{4K \!+\! 2} \right)}}
\label{equ19}
\end{equation}
\end{small}
\begin{small}
\begin{equation}
{{\bf{W}}_3} = {\left[ {\begin{array}{*{20}{c}}
{{{\bf{M}}_3}}&{1/2{B_k}{{\bf{e}}_k}}\\
{1/2{B_k}{\bf{e}}_k^T}&{{0}}
\end{array}} \right]_{\left( {4K + 2} \right) \times \left( {4K + 2} \right)}}
\label{equ20}
\end{equation}
\end{small}
\begin{small}
\begin{equation}
{{\bf{W}}_4} = {\left[ {\begin{array}{*{20}{c}}
{{{\bf{M}}_4}}&{ - 1/2T_{loc,k}{{\bf{e}}_k}}\\
{ - 1/2T_{loc,k}{\bf{e}}_k^T}&0
\end{array}} \right]_{\left( {4K + 2} \right) \times \left( {4K + 2} \right)}}
\label{equ21}
\end{equation}
\end{small}
\begin{small}
\begin{equation}
{{\bf{W}}_5} = {\left[ {\begin{array}{*{20}{c}}
{{{\bf{0}}_{({4K + 1}) \times ({4K + 1}})}}&{1/2{{\bf{e}}_{3K + k}}}\\
{1/2{\bf{e}}_{3K + k}^T}&0
\end{array}} \right]_{\left( {4K + 2} \right) \times \left( {4K + 2} \right)}}.
\label{equ22}
\end{equation}
\end{small}

In the problem (\ref{EqObj4}), only the constraint (\ref{EqObj4H}) is non-convex. By dropping the rank-1 constraint (\ref{EqObj4H}), (\ref{EqObj4}) becomes a positive SDP problem which can be solved using standard CVX tools \cite{MEC15}.

Define ${{\bf{G}}^*}$ as the optimal solution of the problem (\ref{EqObj4}) without the rank-1 constraint, and note that ${\bf{G}}\left( {4K + 2,k} \right) = s\left( k \right)$ for $k = 1, \ldots ,4K + 1$. Therefore, ${{\bf{G}}^*}\left( {4K + 2,k} \right)$, ($k = 1, \ldots ,K$) can be used to recover the binary offloading decision ${c_k}$. Define ${\bf{D}} = {\left[ {{d_1}, \ldots ,{d_K}} \right]^T}$, and let $D = {\left[ {{{\bf{G}}^*}(4K + 2,1), \ldots ,{{\bf{G}}^*}(4K + 2,K)} \right]^T}$. Note that ${d_k} \in \left[ {0,1} \right]$ for $k = 1, \ldots ,K$.  Define ${\bf{\tilde {\bf{c}}}} = {\left[ {{{\tilde c}_1}, \ldots ,{{\tilde c}_K}} \right]^T}$ as the feasible binary offloading decision vector and $\gamma $ as the decision making threshold, if ${d_k} > \gamma $, ${\tilde c_k} = 1$; otherwise, ${\tilde c_k} = 0$. The overall offloading decision making and MU-MIMO Computation Offloading (DM-MMCO) algorithm is summarized in Algorithm 1.
\begin{algorithm}[htb]
\caption{Offloading Decision Making and MU-MIMO Computation Offloading (DM-MMCO)}
\label{alg1}
\begin{algorithmic}[1]
\STATE  Initialization:
system parameters $K,B_W,{\sigma^2},{M},{N}$;\\
parameters of MDs ${B_k},{C_k},f_{loc,k},f_{c,k},P_{\max}^k,p_k^{id},\tau_k^{\max }$,\\
$\forall k \in {\mathds{U}}$; Matrices in the problem (\ref{EqObj4}) ${{\bf{W}}_0},{{\bf{W}}_1},{{\bf{W}}_2},{{\bf{W}}_3},{{\bf{W}}_4},{{\bf{W}}_5},\forall k \in {\mathds{U}}$.\\
\STATE  Solve the problem (\ref{EqObj4}) to obtain optimal solution ${G^*}$.\\
\STATE  Extract the values of ${\bf{G}}\left( {4K + 2,k} \right),k = 1, \ldots ,K$\\
\rm{ }\rm{ }\rm{ }to ${\bf{D}} = {\left[ {{d_1}, \ldots ,{d_K}} \right]^T}$.\\
\STATE  Obtain offloading decision vector ${\bf{\tilde c}}$ according\\
\rm{ }\rm{ }\rm{ }to ${\bf{D}}$ and $\gamma $.\\
\STATE  Solve the MU-MIMO beamforming problem for MDs in ${{\mathds{U}}_o}$ under ${\bf{\tilde {\bf{c}}}}$.
\STATE  Output: Offloading decision vector ${\bf{\tilde {\bf{c}}}}$, beamforming matrices sets ${\mathds{Q,V}}$\\
\end{algorithmic}
\end{algorithm}

\section{MU-MIMO Beamforming Design for Computation Offloading}
\label{section5}
When the offloading decision vector ${\bf{\tilde c}}$ is achieved as stated in section \ref{section4}, the problem (P1) can be rewritten as
\begin{small}
\begin{equation}
P5:{\rm{ }}\mathop {\min }\limits_{{\mathds{Q,V}}} {\rm{ }}\sum\limits_{k \in {{\mathds{U}}_o}} \!\!{\left( \!{\lambda _k^e\frac{{{B_k}}}{{{R_k}}}\left\| {{{\bf{Q}}_k}} \right\|_F^2 \!+\! \lambda _k^t\mathop {max}\limits_{k \in {{\mathds{U}}_o}} \left\{ {\frac{{{B_k}}}{{{R_k}}}} \right\}} \!\!\right)}  \!+\! \zeta {\rm{ }}
\label{equ23}
\end{equation}
\end{small}subject to constraints (\ref{EqObj1C})-(\ref{EqObj1D}), where $\zeta  \!=\! \sum\nolimits_{i \in {{\mathds{U}}_l}} {\left( {\lambda _i^eE_{loc,i} + \lambda _i^tT_{loc,i}} \right)}  + \sum\nolimits_{j \in {{\mathds{U}}_o}} {\left( {\lambda _j^eE_{c,j} + \lambda _j^tT_{c,j}} \right)} $ is a constant and can be omitted in the following analysis. Obviously, the problem (\ref{equ23}) is a nonconvex problem and is difficult to solve. Let ${\max _{k \in {{\mathds{U}}_o}}}\left\{ {{B_k}/{R_k}} \right\} = q$ and the problem (\ref{equ23}) can be transformed into
\begin{small}
\begin{subequations}
\label{EqObj6}
\begin{eqnarray}
& P6:{\rm{  }}\mathop {\min }\limits_{{\mathds{Q,V}},q} {\rm{ }}\sum\limits_{k \in {{\mathds{U}}_o}} {\left( {\lambda _k^e\frac{{{B_k}}}{{{R_k}}}\left\| {{{\bf{Q}}_k}} \right\|_F^2 + \lambda _k^tq} \right)} \label{EqObj6A} \\
& s.t.{\rm{ }}\left\| {{{\bf{Q}}_k}} \right\|_F^2 \le P_{\max }^k{\rm{  }}\forall k \in {{\mathds{U}}_o} \label{EqObj6B} \\
& q + T_{c,k} \le \tau _{\max }^k,\forall k \in {{\mathds{U}}_o} \label{EqObj6C} \\
& \frac{{{B_k}}}{{{R_k}}} \le q,\forall k \in {{\mathds{U}}_o} \label{EqObj6D}.
\end{eqnarray}
\end{subequations}
\end{small}

To solve the problem (P6), the quadratic transform proposed in \cite{MEC16} is adopted to solve the problem (\ref{EqObj6}) in this paper. Then, a new objective is given by ${f_q}\left( {{\mathds{Q,V,W}},q} \right) = \sum\limits_{k \in {{\mathds{U}}_o}} {\left( {2{w_k}\sqrt {\lambda _k^e{B_k}\left\| {{{\bf{Q}}_k}} \right\|_F^2}  - w_k^2{R_k} + \lambda _k^tq} \right)}$, where ${\mathds{W}}=\left\{ {{w_k},k = 1, \ldots ,{N_o}} \right\}$ with ${w_k} \in {\mathds{R}}$ introduced for each offloading MD. The the optimization problem (\ref{EqObj6}) can now be recast to
\begin{small}
\begin{subequations}
\label{EqObj7}
\begin{eqnarray}
& P7:{\rm{ }}\mathop {\min }\limits_{{\mathds{Q,V,W}},q} {\rm{ }}{f_q}\left( {{\mathds{Q,V,W}},q} \right) \label{EqObj7A} \\
& s.t.~~{\rm{  }}\left\| {{{\bf{Q}}_k}} \right\|_F^2 - P_{\max }^k \le {\rm{ 0, }}\forall k \in {{\mathds{U}}_o} \label{EqObj7B} \\
& q + T_{c,k} - \tau _{\max }^k \le 0,\forall k \in {{\mathds{U}}_o} \label{EqObj7C} \\
& \frac{{{B_k}}}{{{R_k}}} - q \le 0,\forall k \in {{\mathds{U}}_o} \label{EqObj7D}.
\end{eqnarray}
\end{subequations}
\end{small}

Then, the multidimensional quadratic transform in \cite{MEC16} is applied to each SINR term inside the ${R_k}$ expression in (\ref{equ3}), and further recast ${f_q}$ to ${f_{qm}}$ as in (\ref{equ26})
\begin{small}
\begin{equation}
{f_{qm}}({\mathds{Q,\!V,\!W,\!Z}},q) \!\!=\!\! \!\!\sum\limits_{k \in {{\mathds{U}}_o}} \!\!\!{\left( \!\!{2{w_k}\sqrt {\lambda _k^e{B_k}\left\| {{{\bf{Q}}_k}} \right\|_F^2}  \!+\! \lambda _k^tq \!-\! \xi } \!\!\right)}
\label{equ26}
\end{equation}
\end{small}where $\xi $ is defined as

$\xi {\rm{ = }}{\!\!}\sum\nolimits_{l = 1}^d {\!\!}{w_k^2B_{W}{{\log }_2}{\!\!}\left( {\!}{1 {\!\!}+{\!\!} 2{\mathop{\rm Re}\nolimits} {\!\!}\left\{ {\!\!}{z_{k(l)}^Hv_{k(l)}^H{H_k}{q_{k(l)}}} {\!\!}\right\} {\!\!}-{\!\!} z_{k(l)}^H{I_N}{z_{k(l)}}} {\!\!}\right)}$.

The final reformulation of the problem (\ref{EqObj6}) after the twice use of the quadratic transform now becomes
\begin{small}
\begin{subequations}
\label{EqObj8}
\begin{eqnarray}
& P8:{\rm{  }}\mathop {\min }\limits_{{\mathds{Q,V,W,Z}},q} {\rm{ }}{f_{qm}}\left( {{\mathds{Q,V,W,Z}},q} \right) \label{EqObj8A} \\
& s.t.~~{\rm{  }}\left\| {{{\bf{Q}}_k}} \right\|_F^2 - P_{\max }^k \le {\rm{ 0, }}\forall k \in {{\mathds{U}}_o} \label{EqObj8B} \\
& q + T_{c,k} - \tau _{\max }^k \le 0,\forall k \in {{\mathds{U}}_o} \label{EqObj8C} \\
& \frac{{{B_k}}}{{{R_k}}} - q \le 0,\forall k \in {{\mathds{U}}_o} \label{EqObj8D}
\end{eqnarray}
\end{subequations}
\end{small}where ${\mathds{Z}}$ denotes the set of auxiliary variables $\left\{ {{z_{k(l)}} \in C|\forall k \in {{\mathds{U}}_o},l = 1, \ldots ,d} \right\}$ with ${z_{k(l)}} \in {\mathds{C}}$ introduced for each data stream $\left( {k,l} \right)$. Based on the above definitions, we can see that when ${\mathds{Z}}$ and ${\mathds{W}}$ are both fixed, the problem (\ref{EqObj8}) is a convex problem over ${\mathds{Q}}$ and ${\mathds{V}}$, so the optimal ${\mathds{Q}}$ and ${\mathds{V}}$ can be efficiently found by using the standard numerical method \cite{MEC17}.

Therefore, we propose an iterative optimization algorithm to solve the problem (\ref{EqObj8}). According to \cite{MEC16}, when all the other variables are fixed, the optimal ${z_{k(l)}}$ is given by
\begin{small}
\begin{equation}
 z_{k(l)}^* = I_N^{ - 1}v_{k(l)}^H{H_k}{q_{k(l)}}.
 \label{equ28}
\end{equation}
\end{small}

After updating ${z_{k(l)}}$, the optimal ${w_k}$ is
\begin{small}
\begin{equation}
w_k^* = \frac{{\sqrt {\lambda _k^e{B_k}\left\| {{{\bf{Q}}_k}} \right\|_F^2} }}{{{R_k}}}.
\label{equ29}
\end{equation}
\end{small}

Then, the optimal ${\mathds{Q}}$ is obtained by solving the problem (\ref{EqObj8}) when ${\mathds{V}}$, ${\mathds{Z}}$ and ${\mathds{W}}$ are fixed. At last, the optimal ${\mathds{V}}$ is achieved by solving the problem (\ref{EqObj8}) when ${\mathds{Q}}$, ${\mathds{Z}}$ and ${\mathds{W}}$ are fixed. The whole iterative optimization algorithm for solving (\ref{EqObj8}) is summarized in Algorithm 2.
\begin{algorithm}[htb]
\caption{MU-MIMO Beamforming Design for MU-MIMO Computation Offloading}
\label{alg2}
\begin{algorithmic}[1]
\STATE  Initialization:
${\bf{Q}}_k^{(0)},{\bf{V}}_k^{(0)},(\forall k \in {{\mathds{U}}_o})$ into feasible values;\\
\rm{  }\rm{  }\rm{  }iteration number: $n = 1$; maximum number of iterations: $numiter$; $f_{qm}^{(0)} = 0$ and tolerance $\varepsilon $.
\STATE  for $n = 1$ to $numiter$
\STATE  Update auxiliary variables' set ${\mathds{Z}}$ and ${\mathds{W}}$ with fixed ${\bf{Q}}_k^{(n - 1)},{\bf{V}}_k^{(n - 1)},(\forall k \in {{\mathds{U}}_o})$ by (\ref{equ28}) and (\ref{equ29}). Then\\
  \rm{  }\rm{  }\rm{  }optimize ${\bf{Q}}_k^{(n)},(\forall k \in {{\mathds{U}}_o})$ with fixed ${\mathds{Z}}$, ${\mathds{W}}$ and ${\bf{V}}_k^{(n - 1)},(\forall k \in {{\mathds{U}}_o})$ by solving the optimization problem (\ref{EqObj8}).\\
\STATE  Update auxiliary variables' set ${\mathds{Z}}$ and ${\mathds{W}}$ with fixed\\
 \rm{  }\rm{  }\rm{  }${\bf{Q}}_k^{(n)},{\bf{V}}_k^{(n - 1)},(\forall k \in {{\mathds{U}}_o})$ by (\ref{equ28}) and (\ref{equ29}). Then optimize\\
 \rm{  }\rm{  }\rm{  }${\bf{V}}_k^{(n)},\forall k \in {{\mathds{U}}_o}$ with fixed ${\mathds{Z}}$, ${\mathds{W}}$ and ${\bf{Q}}_k^{(n)},(\forall k \in {{\mathds{U}}_o})$\\
   \rm{  }\rm{  }\rm{  }by solving the optimization problem (\ref{EqObj8}).\\

\STATE  Calculate $\left| {f_{qm}^{(n)} - f_{qm}^{(n - 1)}} \right|$, if $n > numiter$\\
 \rm{  }\rm{  }\rm{  }or $\left| {f_{qm}^{(n)} - f_{qm}^{(n - 1)}} \right| < \varepsilon $, terminate.\\
\STATE  end for.\\
\STATE  Output: MU-MIMO beamforming matrices set ${\mathds{Q}}$ and ${\mathds{V}}$,\\
and $\min \left\{ {f_{qm}^{(i)}|i = 1, \ldots ,n} \right\}$.\\
\end{algorithmic}
\end{algorithm}

\section{Simulation Results}
\label{section5}
In this section, simulation results are provided to show the performance of the proposed algorithms. Simulations is performed on a Monte Carlo simulation on a Matlab-based simulator. We adopt the 3GPP pathloss model \cite{MEC18} and Rayleigh fading with zero mean and unit variance. The background noise density is set to be $-175$dBm/Hz and $B_W=10$MHz. ${B_k}$ is uniformly chosen from 0.8MB to 1.2MB. We have $P_{\max }^k{\rm{ = 0}}{\rm{.1W}}$, $\tau _{\max }^k{\rm{ = 3s}}$ and $p_k^{id} {\rm{ = 0}}{\rm{.005W}}$ for each MD. ${f_{loc,k}}$ and ${f_{c,k}}$ for MD $k$ are uniformly chosen from 0.2G to 0.5G cycles/s and 0.8G to 1G cycles/s, respectively. Unless stated otherwise, some system parameters are set as follows: ${M}{\rm{ = }}16$, ${N} = d = 2$, $\kappa  = {10^{ - 25}}$, $\alpha  = 237.5$ cycles/bit, $\gamma {\rm{ = }}0.8$, and error tolerance $\varepsilon {\rm{ = }}{10^{ - 3}}$. Note that we set $\lambda _k^e = 1$ and $\lambda _k^t = 0$ as the default values. For comparison, we also simulate the following offloading schemes: 1) Local-only: all the MDs compute their tasks locally; 2) OP-MMSE: each MD uses single antenna to offload computation tasks to the BS simultaneously. The MDs only have transmit power control, and MMSE receiver is used at BS side; 3) FDMA: MDs equipped with single antenna use orthogonal frequency resources to offload computation tasks and maximum transmit power is used; 4) TDMA: MDs equipped with single antenna sequentially offload computation tasks to the BS and maximum transmit power is used.

\begin{figure}
\centering
\includegraphics[width=8cm]{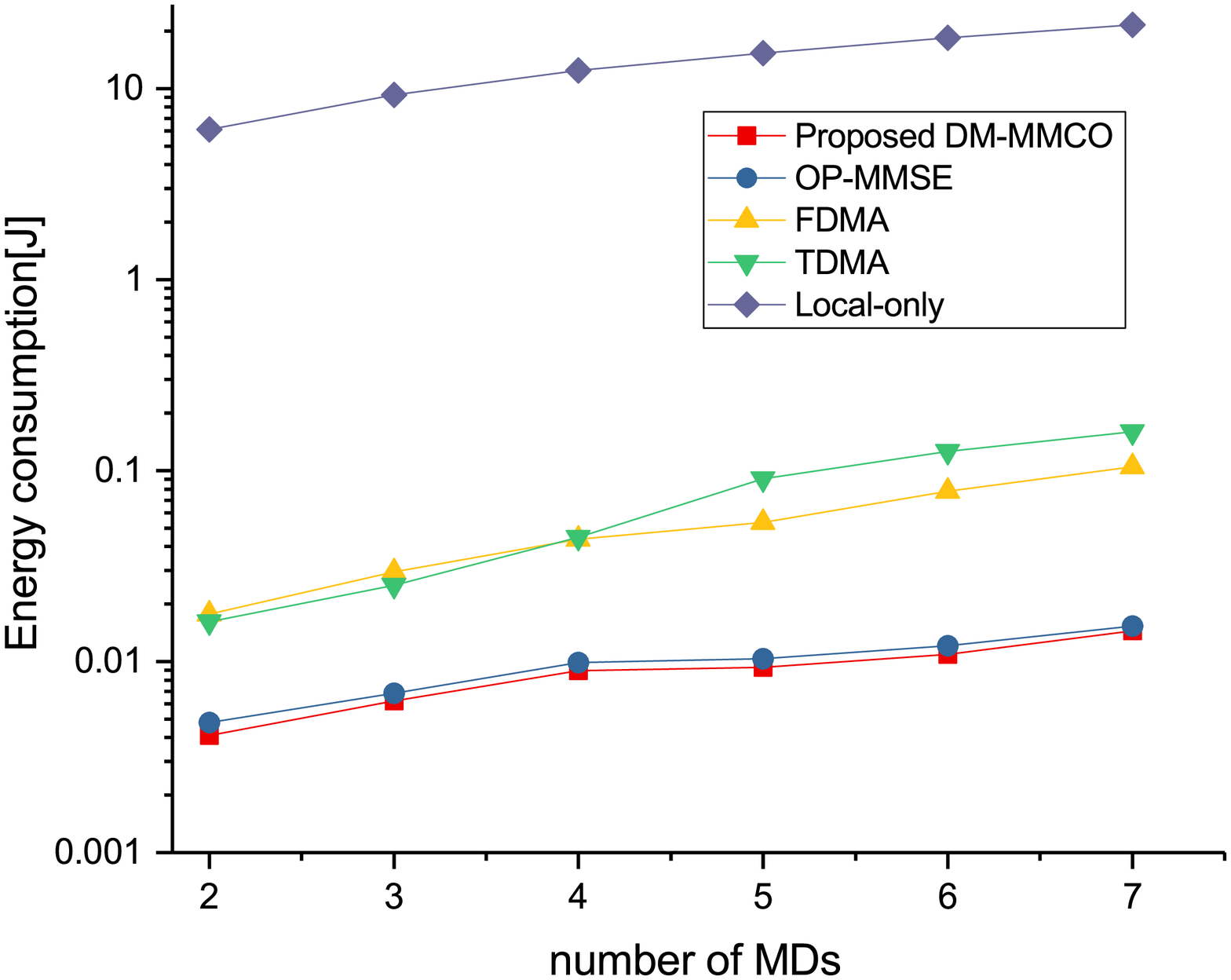}
\caption{Energy consumption versus the number of MDs}
\label{fig2}
\end{figure}

Figure \ref{fig2} shows the energy consumption versus the number of MDs for different offloading schemes. It can be observed that the energy consumption of all offloading schemes increases with the increasing number of MDs. From the figure, it can be seen that the proposed DM-MMCO has the lowest energy consumption. It is because that the transmission time delay can be greatly reduced by using MIMO transmission and the energy efficiency is improved. Since the MD has limited computation capability, the local-only method has the highest energy consumption.
It can be shown that without transmit beamforming and multiple data streams, the performance of OP-MMSE is inferior than the proposed DM-MMCO. The energy consumption for FDMA and TDMA grows quickly when the number of MDs increases. It can also be seen that when the number of MDs is small (e.g. 2, 3 and 4), the energy consumption of TDMA is smaller than FDMA. However, when the number of MDs is larger than 4, the energy consumption of TDMA grows fastly and is obviously larger than the energy consumption of FDMA. The reason is that the time delay will accumulate in TDMA and some MDs will choose to compute tasks at local if the maximum tolerable delay is not satisfied. This leads to the increasing of energy consumption in TDMA. It can be noted that the performance gap between the proposed DM-MMCO and OP-MMSE is small, the reason is that the power control method in OP-MMSE also adopts the quadratic transform proposed in \cite{MEC16} and can mitigate inter-user interference at certain degree. In addition, the quadratic transform proposed in \cite{MEC16} is similar to the WMMSE algorithm, which is described in \cite{MEC19}. Therefore, the performance gap between the proposed DM-MMCO and OP-MMSE is small. However, the proposed FP framework in \cite{MEC16} is more computationally efficient than WMMSE, which is proved in \cite{MEC19}.

\begin{figure}
\centering
\includegraphics[width=8cm]{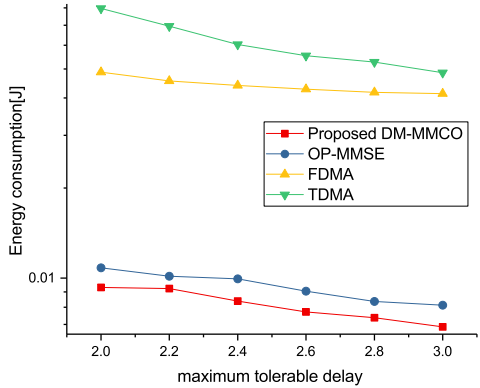}
\caption{Average time delay versus the number of MDs}
\label{fig4}
\end{figure}

Figure \ref{fig4} shows the impact of maximum tolerable delay on the energy consumption of different offloading schemes. It can be observed that with the increasing of maximum tolerable delay, all offloading schemes experience the decreasing of energy consumption. It can also be seen that the energy consumption of TDMA drops quickly with the increasing of maximum tolerable delay, which means more MDs choose to offload computation tasks and more energy consumption is reduced. It is shown that the proposed DM-MMCO has the lowest energy consumption among all the offloading schemes. It should be noted that the energy consumption of both the OP-MMSE scheme and the proposed DM-MMCO scheme decreases quickly with the increasing of maximum tolerable delay. The reason is that with the increasing of maximum tolerable delay, MDs in DM-MMCO or OP-MMSE can use less transmission power to finish task offloading, thus the energy consumption is reduced.

\section{Conclusions}
\label{section7}
In this paper, a joint computation offloading and MU-MIMO transmission problem was studied in a MEC system. A joint optimization of offloading decision-making and MU-MIMO beamforming problem was formulated to minimize MDs' cost under maximum tolerable delay and transmission power constraints. To solve the optimization problem, two low complexity algorithms were proposed to obtain the offloading decisions and MU-MIMO beamforming matrices, respectively. Simulation results showed that the proposed algorithms have excellent performance in reducing the energy consumption and time delay of computation offloading.

\section{Acknowledgements}
This work is supported in part by the National Nature Science Foundation of China under Grant 61571115 and 61701254, and in part by Natural Science Foundation of Jiangsu Province BK20170901.

\end{document}